\documentclass{applemlr}

\usepackage{amsmath}
\usepackage{enumerate}
\usepackage{algorithm}
\usepackage{algpseudocode}
\usepackage{amsfonts}
\usepackage{amsthm}
\usepackage{cleveref}
\usepackage{diagbox}
\usepackage{colortbl}
\usepackage{amssymb}
\usepackage{xspace}
\usepackage{wrapfig}
\usepackage{adjustbox}
\usepackage{tabularx}
\usepackage{booktabs}
\usepackage{mathtools}
\usepackage{tikz}
\usepackage{enumitem}
\usepackage{silence}
\usepackage{dsfont}
\usepackage{multirow}
\usepackage{makecell}
\usepackage{xfakebold}

\usepackage{amsmath,amsfonts,bm}

\def\eqref#1{equation~\ref{#1}}

\def\1{\bm{1}}

\DeclareMathAlphabet{\mathsfit}{\encodingdefault}{\sfdefault}{m}{sl}
\SetMathAlphabet{\mathsfit}{bold}{\encodingdefault}{\sfdefault}{bx}{n}

\definecolor{textgray}{HTML}{6E6E73}
\usetikzlibrary{positioning, calc}
\usetikzlibrary{decorations.pathmorphing}

\makeatletter
\patchcmd{\wrong@fontshape}{\@gobbletwo}{}{}{}
\makeatother
\numberwithin{equation}{section}
\makeatletter
\AtBeginDocument{
  \urlstyle{sf}
  
}
\makeatother

\definecolor{light}{RGB}{125, 125, 125}
\crefname{tcb@cnt@pbox}{code}{code}
\Crefname{tcb@cnt@pbox}{Code}{Code}
\crefname{assumption}{assumption}{assumption}
\Crefname{assumption}{Assumption}{Assumptions}

\newtcolorbox[auto counter]{pbox}[2][]{
  colback=white,
  title=Code~\thetcbcounter: #2,
  #1,fonttitle=\sffamily,
  fontupper=\sffamily,
  arc=2pt,
  colframe=bgcolor,
  coltitle=fgcolor,
  colbacktitle=bgcolor,
  toptitle=0.25cm,
  bottomtitle=0.125cm
}

\makeatletter
\newcommand\applefootnote[1]{%
  \begingroup
  \renewcommand\thefootnote{}%
  \renewcommand\@makefntext[1]{\noindent##1}%
  \footnote{#1}%
  \addtocounter{footnote}{-1}%
  \endgroup
}
\makeatother

\definecolor{cverbbg}{gray}{0.90}

\title{Which Data Matter? Embedding-Based Data Selection for Speech Recognition}

\author{Zakaria Aldeneh$^1$}
\author{Skyler Seto$^1$}
\author{Maureen de Seyssel$^1$}
\author{Jie Chi$^1$}
\author{Zijin Gu$^1$}
\author{Takuya Higuchi$^1$}
\author{Jee-weon Jung$^1$}
\author{\\Shinji Watanabe$^2$}
\author{David Grangier$^1$}
\author{Barry-John Theobald$^1$}
\author{Tatiana Likhomanenko$^1$}

\affiliation{$^1$Apple}
\affiliation{$^2$Carnegie Mellon University}

\abstract{Modern ASR systems are typically trained on large-scale pseudo-labeled, in-the-wild data spanning multiple domains. While such heterogeneous data benefit generalist models designed for broad deployment, they pose challenges for specialist models targeting specific domains: specialist models lack the capacity to learn from all available data, and one must pay closer attention to addressing the mismatch between training and test conditions. In this work, we study targeted data selection as a strategy to address these challenges, selecting relevant subsets from $100$k hours of in-the-wild training data to optimize performance on target domains. We represent speech samples using embeddings that capture complementary characteristics---speaker attributes, phonetic content, and semantic meaning---and analyze how relevance and diversity along these axes when performing data selection affect downstream ASR performance. Our experiments with CTC-based Conformer models show that training on a strategically selected $5\%$ subset can exceed the performance of models trained on the full dataset by up to $36.8\%$ relative WER reduction on target domains.}

\metadata[Correspondence]{\sffamily Zakaria Aldeneh: \url{zaldeneh@apple.com}; Tatiana Likhomanenko: \url{antares@apple.com}}
\date{\sffamily\today}

\begin{document}

\maketitle

\section{Introduction}
The curation of large-scale training datasets has had a notable impact on speech processing tasks, including automatic speech recognition (ASR)~\citep{ardila2020common,chan2021speechstew,wang2021voxpopuli,radford2023robust}. Such datasets expose speech models to a wide range of acoustic conditions, speakers, and recording scenarios, enabling them to learn robust representations and achieve strong performance. However, such success is typically achieved with generalist models that have the capacity to leverage the scale and heterogeneity of the training data. In practice, we often deploy specialist models tailored to specific domains or tasks---models that lack the capacity to learn from all available training data effectively. For these specialist models, domain mismatch between training and test conditions becomes essential: a specialist model cannot capture all the details of a large, heterogeneous dataset and must instead focus on data closely aligned with the target conditions. This scenario motivates our central question: \textit{Can we strategically select subsets of large-scale, in-the-wild data that enable specialist models to outperform models trained on the full dataset?}

Mismatch between the train and test domains along any dimension of variation in speech can degrade performance. Prior work has shown, for example, that models trained primarily on read speech perform poorly when evaluated on spontaneous speech~\citep{likhomanenko2020rethinking}. Similarly, the performance of models trained on native accents degrades when applied to accented or non-native speech~\citep{fukuda2018data,ghorbani2019leveraging,zhang2022mitigating}. These observations highlight that not all training data are equally useful for a given target domain, and an effective data selection strategy cannot rely on a single notion of similarity. Effective data selection, therefore, requires capturing relevance across multiple, complementary axes of speech variability, rather than relying on a single notion of similarity when addressing the mismatch between train and test conditions.

We study targeted data selection for ASR in the regime of large-scale in-the-wild training data (over $100$k hours) and specialist models with approximately $10$-$100$M parameters (production-size models). We represent speech samples using embeddings that capture diverse properties---speaker, phonetic, and semantic---and use them to select subsets of data that are both relevant to the target domain and diverse along these dimensions. We analyze which characteristics matter most for downstream performance, and whether combining multiple representations yields additional gains. Our experiments span multiple target datasets and operate at a scale representative of modern ASR training pipelines that rely on large-scale in-the-wild pseudo-labeled data, offering insights into effective data selection in that regime. Specifically, we show that one can achieve up to $36.8\%$ improvement in performance on target domains with data selection in this training regime, even with $5\%$ of the total available training data.

\section{Related Work}
Data selection has been studied in the context of ASR~\citep{wu2007data,wei2014unsupervised,wei2014submodular,park2022unsupervised,zhang2006new,lin09e_interspeech,zheng2023unsupervised,doulaty2015data,afshan2021sequence,lu2022unsupervised,kothawade2023ditto}. Prior work approached the problem from several perspectives: improving training efficiency by reducing dataset size~\citep{wu2007data,wei2014unsupervised,wei2014submodular,park2022unsupervised}, selecting samples for active learning~\citep{zhang2006new,zheng2023unsupervised}, selecting samples to annotate by humans under limited annotation budgets~\citep{lin09e_interspeech,nallasamy2012active}, and improving domain-specific performance by prioritizing relevant data~\citep{doulaty2015data,afshan2021sequence,lu2022unsupervised,kothawade2023ditto}. Our work focuses on the last setting: selecting relevant samples from large-scale in-the-wild training corpora to accommodate train/test mismatches to improve performance on a specific target domain.

Confidence-based strategies have been widely used in semi-supervised and active learning settings, where samples are chosen based on model prediction confidence. However, selecting data solely based on confidence has been shown to be suboptimal for downstream performance~\citep{zhang2006new}. These findings motivated distribution-matching approaches that account for domain mismatch. For example, \citet{wu2007data} selected subsets with uniform coverage over linguistic units such as words or phonemes.
Other methods used learned representations for data selection. Acoustic-based approaches have used i-vectors~\citep{siohan2013ivector} and other low-level signal statistics to measure similarity between candidate and target samples. \citet{lu2022unsupervised} proposed unsupervised selection using discrete speech units derived from self-supervised models. \citet{afshan2021sequence} used sequence-level confidence scores from pre-trained ASR models to identify informative samples for domain adaptation. \citet{rangappa2025speech} combined domain classifiers with pseudo-label filtering to select unlabeled data for fine-tuning. In the context of self-supervised learning, \citet{gody2023unsupervised} explored data selection for fine-tuning pre-trained speech models on downstream tasks.

We differ from prior work in three key aspects: (1) \textit{Scale}: we operate on $100$k+ hours of in-the-wild pseudo-labeled data from Granary~\citep{li2023yodas,rao2025granary} to train production-scale specialist models (10-100M parameters), whereas most prior work used small-scale datasets (e.g., Switchboard, LibriSpeech); (2) \textit{Multi-embedding selection}: we capture complementary speech characteristics---speaker, phonetic, and semantic---and analyze both their individual impact and the benefits of combining them, rather than focusing on a single aspect; (3) \textit{Multi-target selection}: we balance relevance and diversity while selecting for multiple target domains simultaneously, rather than optimizing for one domain at a time. \textbf{Overall, we study a setting increasingly common in practice: using large-scale, diverse in-the-wild pseudo-labeled data to train specialist ASR models for specific target domains.}

\section{Approach}
\subsection{Subset Selection}

Let $D_{\text{source}}$ denote a large-scale in-the-wild training dataset and $D_{\text{target}}$ a small validation set drawn from the target domain. Note that $D_{\text{target}}$ is small (e.g., one hour of speech) and insufficient for training a model from scratch. We seek to select a subset $\mathcal{S} \subset D_{\text{source}}$, substantially smaller than the full dataset, that is informed by $D_{\text{target}}$ and can be used to train an ASR model that performs well on the target domain. Let $\phi$ be an embedding extractor that maps a variable-length utterance $u_i$ to a fixed-size embedding $x_i\in \mathbb{R}^d$, capturing relevant characteristics of the utterance. For example, if $\phi$ is a speaker embedding function, similarity between embeddings reflects closeness in speaker characteristics. Using these embeddings, we can identify source samples that are most relevant to the target domain. Specifically, for each candidate training sample $x \in \mathcal{D}_{\text{source}}$, we define its relevance to the target set as the maximum similarity to any target embedding:
\begin{align}
\text{sim}(x, \mathcal{D}_{\text{target}}) = \max_{y \in \mathcal{D}_{\text{target}}} \text{sim}(x, y),
\end{align}
where $\text{sim}(\cdot,\cdot)$ is a similarity measure, e.g., cosine similarity. 

A simple strategy would be to select the top-$k$ samples by this relevance score. However, this strategy can lead to redundant samples that provide overlapping information. To address this limitation, we use Maximal Marginal Relevance (MMR)~\citep{carbonell1998use}, which balances relevance to the target set with diversity among the selected samples in an \textit{iterative} fashion. At each selection step, MMR chooses the next sample $x$ by maximizing:
\begin{equation}
\label{eq2}
\text{MMR}(x) = \lambda \, \text{sim}(x, \mathcal{D}_{\text{target}}) - (1-\lambda) \, \max_{s \in S} \text{sim}(x, s),
\end{equation}
where $S$ is the set of samples already selected, and $\lambda \in [0,1]$ controls the trade-off between relevance and diversity. This procedure produces a subset that is both representative of the target domain and non-redundant, improving model generalization compared to random selection.

\subsection{Embedding Selection}
Our data selection framework relies on an embedding function $\phi(\cdot)$ that captures relevant characteristics of speech to quantify the similarity between source and target utterances. A key design choice is therefore which characteristics of the speech signal should be emphasized when defining relevance and diversity. We evaluate three embeddings---\emph{speaker embeddings}~\citep{jung2024espnet}, \emph{WavLM embeddings}~\citep{chen2022wavlm}, and \emph{SBERT embeddings}~\citep{reimers-2019-sentence-bert}---each capturing distinct aspects of speech. By comparing their effectiveness within the same MMR-based selection framework, we can determine which characteristics are most important for data selection in ASR training and separate the benefits of embedding-based selection from other factors.

\vspace{-0.3cm}
\paragraph*{Speaker Embeddings (Speaker Characteristics).}
Speaker embeddings are designed to encode speaker-specific attributes such as vocal tract characteristics, demographic cues, and speaking style, while being largely invariant to linguistic content. However, prior probing studies~\citep{raj2019probing} have shown that speaker embeddings capture additional factors beyond speaker identity. These factors include session conditions, recording environments, data augmentations, and even transcription-related artifacts. As a result, using speaker embeddings for data selection biases the selected subset toward source samples whose speakers and acoustic conditions are most similar to those in the target domain. 

\vspace{-0.3cm}
\paragraph*{WavLM Embeddings (Phonetic Characteristics).}
WavLM embeddings are designed to encode phonetic and sub-phonetic information while being largely invariant to speaker identity and acoustic variations~\citep{chen2022wavlm}. WavLM extends HuBERT~\citep{hsu2021hubert} by incorporating noise-robust and mixture-aware pretraining objectives, enabling it to learn phonetic representations that are invariant to background noise, reverberation, and speaker overlap. Prior probing studies on HuBERT have shown that its representations strongly correlate with linguistic units such as phones and articulatory features~\citep{wells2022phonetic,choi24b_interspeech,sicherman2023analysing}. As a result, using WavLM embeddings for data selection biases the selected subset toward source samples that are similar to the target domain in terms of phonetic coverage and pronunciation patterns rather than speaker characteristics. 

\vspace{-0.3cm}
\paragraph*{SBERT Embeddings (Semantic Characteristics).}
SBERT embeddings~\citep{reimers-2019-sentence-bert} are derived from textual transcripts and capture the semantic and syntactic properties of utterances. Using SBERT for data selection emphasizes coverage of linguistic meaning, vocabulary, and sentence structure present in the target domain. Using SBERT embeddings enables us to assess whether aligning the semantic distributions of the training data with those of the target domain improves ASR performance, especially in domains with distinctive topics or lexical distributions. Unlike speaker or WavLM embeddings, SBERT embeddings require access to transcripts.

\subsection{Multi-Embedding Subset Selection}
\label{subsec:multi-embedding}
In the single-embedding setting, MMR selects samples based on a trade-off between relevance to the target domain and diversity among selected samples, using a single embedding type, $\phi(\cdot)$. To use multiple embedding types, we can generalize the relevance and diversity computations as follows.
Let each utterance $u_i$ have $K$ embeddings: \[x_i^{(1)}, x_i^{(2)}, \dots, x_i^{(K)}; \quad x_i^{(k)} \in \mathbb{R}^{d_k},\]
where each embedding corresponds to a different captured characteristic of speech (e.g., speaker, phonetic, or semantic).
For relevance to the target domain, we compute a similarity score for each embedding type and combine them via a weighted sum:
\begin{align}
\text{sim}_\text{multi}(x_i, \mathcal{D}_\text{target}) = \sum_{k=1}^K \,w_k \max_{y \in \mathcal{D}_\text{target}^{(k)}} \text{sim}\big(x_i^{(k)}, y^{(k)}\big),
\end{align}
where $w_k$ controls the contribution of the $k$-th embedding.

Similarly, to encourage diversity, we compute for each embedding the maximum similarity between the candidate utterance and the already selected set $S$, and aggregate across embeddings using a weighted sum:
\begin{align}
v_i = \sum_{k=1}^K \, w_k\max_{s \in S} \text{sim}\big(x_i^{(k)}, s^{(k)}\big).
\end{align}
The generalized MMR selection criterion becomes:
\begin{align}
\text{MMR}_\text{multi}(x_i) = \lambda \, \text{sim}_\text{multi}(x_i, \mathcal{D}_\text{target}) - (1-\lambda) \, v_i,
\end{align}
where $\lambda \in [0,1]$ again controls the relevance--diversity trade-off. 

This formulation corresponds to a late-fusion strategy, where relevance and diversity are computed separately within each embedding space and combined at the score level. As a result, samples that jointly satisfy multiple embedding criteria are favored. In particular, a candidate that is close to the target domain across all $K$ embeddings receives a high relevance score, while a candidate that is dissimilar to previously selected samples across all embeddings contributes to diversity and is therefore more likely to be selected. While an early-fusion alternative (e.g., concatenating embeddings prior to selection) is possible, late fusion offers greater interpretability and control by allowing explicit weighting of each embedding type via $w_k$.

\subsection{Multi-Dataset Subset Selection}
\label{subsec:multi_data} 
In some settings, we aim for a model that performs well on multiple target domains, without training a model for each domain independently. In these settings, the target domain is composed of multiple  datasets,
$\mathcal{D}_{\text{target}} = \{\mathcal{D}_{\text{target}}^{(1)}, \ldots,
\mathcal{D}_{\text{target}}^{(M)}\}$, each corresponding to a different domain. To measure the relevance of a source sample with embedding $x$ to multiple target datasets, we consider two aggregation strategies: maximum aggregation and mean aggregation.

In the maximum aggregation strategy, a source sample is considered relevant if it closely matches at least one of the
target datasets:
\begin{equation}
\text{sim}_{\max}(x, \mathcal{D}_{\text{target}})
= \max_{m \in \{1,\ldots,M\}}
\max_{y \in \mathcal{D}_{\text{target}}^{(m)}} \text{sim}(x,y).
\end{equation}
This formulation is equivalent to taking the union of the target data and then computing the maximum similarity between the target data and samples from that union. In the mean aggregation strategy, we define the relevance score by averaging the maximum similarity within each target dataset:
\begin{equation}
\text{sim}_{\text{avg}}(x, \mathcal{D}_{\text{target}})
= \frac{1}{M} \sum_{m=1}^M
\max_{y \in \mathcal{D}_{\text{target}}^{(m)}} \text{sim}(x,y).
\end{equation}
This formulation promotes selecting source samples that are useful across multiple target domains rather than overly specialized to a single domain. The resulting relevance score is used within the MMR objective to balance relevance and diversity during subset selection.

\section{Datasets}
We use the English portion of the Granary dataset as our primary source of training data~\citep{rao2025granary}. Granary is a large-scale, open-source, in-the-wild pseudo-labeled collection of speech recordings covering a wide range of speakers and acoustic conditions. Granary is based on YODAS~\citep{li2023yodas} and provides a realistic regime that closely reflects the data conditions encountered when training speech models at scale, where large quantities of pseudo-labeled data must be used to construct effective training sets. Its large scale and diversity ensure that the subset selection process provides sufficient coverage to demonstrate effectiveness across different target domains. Although pseudo-labels may introduce some noise, the dataset’s scale and variability help mitigate this effect, and our experiments focus on measuring relative improvements through subset selection.

To evaluate our subset selection approach, we consider three widely used target speech corpora:

\begin{itemize}
    \item LibriSpeech~\citep{panayotov2015librispeech}: A high-quality dataset derived from audiobooks, commonly used for ASR benchmarking. LibriSpeech contains both clean and noisy recordings across a diverse set of speakers.
    \item CommonVoice English~\citep{ardila2020common}: A crowdsourced dataset with recordings from speakers with varied accents and recording conditions. CommonVoice is useful for evaluating generalization to real-world scenarios.
    \item TED-LIUM~\citep{hernandez2018ted}: A dataset of TED talk recordings, which introduces domain-specific challenges such as spontaneous speech, variable speaking rates, and diverse topics.
\end{itemize}
These target datasets span a broad spectrum of domain characteristics and provide a diverse testbed for our data selection framework. 
Detailed statistics for all datasets are provided in Table~\ref{tab:datasets}.






\begin{table}[t!]
\centering
\caption{Summary of source and target datasets used in our experiments. Granary serves as the large-scale, in-the-wild, training source data. The target datasets---LibriSpeech, CommonVoice, and TED-LIUM---cover a wide range of domains. Dataset sizes (in hours) are reported when available.}
\label{tab:datasets}
\begin{tabular}{lrrrr}
\toprule
\textbf{Type} & \textbf{Dataset} & \textbf{Train (h)} & \textbf{Dev (h)} & \textbf{Test (h)} \\
\midrule
Source & Granary English & 102{,}458 & {--}\phantom{0} & {--}\phantom{0} \\
Target & LibriSpeech        &    961 & 10.5 & 10.7 \\
Target & CommonVoice English        &   1{,}594 & 27.1 & 26.9 \\
Target & TED-LIUM           &    452 &  1.6 &  2.6 \\
\bottomrule
\end{tabular}
\end{table}

\section{Experiments}

\begin{algorithm*}[t]
\caption{Batched Greedy MMR Subset Selection}
\label{alg:greedy-mmr}
\begin{algorithmic}[1]
\Statex\hspace{-\algorithmicindent}\textbf{Input:} Candidate embeddings $X \in \mathbb{R}^{N \times D}$, candidate durations $d \in \mathbb{R}^{N}$, target embeddings $Y \in \mathbb{R}^{M \times D}$, trade-off parameter $\lambda \in [0,1]$, subset fraction $\alpha$, prefilter fraction $\rho$, batch size $B$.
\Statex\hspace{-\algorithmicindent}\textbf{Output:} Selected index set $\mathcal{S}$.

\State $T \gets \alpha \sum_{i=1}^{N} d_i$ \Comment{Target total duration}
\State Compute relevance scores $r_i = \max_{y \in Y} \text{sim}(X_i, y)$ for all~$i$
\State Let $\mathcal{I} \gets$ indices of the top $\rho N$ candidates ranked by $r_i$ \Comment{Relevance prefilter}
\State Initialize $\mathcal{S} \gets \emptyset$, $D_{\text{sel}} \gets 0$

\State Select $i^\ast = \arg\max_{i \in \mathcal{I}} r_i$
\State $\mathcal{S} \gets \mathcal{S} \cup \{i^\ast\}$, \quad $D_{\text{sel}} \gets d_{i^\ast}$

\While{$D_{\text{sel}} < T$}
    \State $\mathcal{R} \gets \mathcal{I} \setminus \mathcal{S}$ \Comment{Remaining candidates}
    \If{$\mathcal{R} = \emptyset$} \textbf{break} \EndIf
    \State Compute diversity scores $v_i = \max_{s \in \mathcal{S}} \text{sim}(X_i, X_s)$ for all $i \in \mathcal{R}$
    \State Compute MMR scores $m_i = \lambda r_i - (1-\lambda) v_i$
    \State Select top-$B$ indices from $\mathcal{R}$ ranked by $m_i$
    \State $\mathcal{S} \gets \mathcal{S} \cup \text{top-}B$, \quad update $D_{\text{sel}}$
\EndWhile

\State \Return $\mathcal{S}$
\end{algorithmic}
\end{algorithm*}

\subsection{Main Experiments}
The goal of the first set of experiments is to assess whether training on chosen subsets aimed at matching the target domain distribution outperforms training on the full, large-scale training corpus.

\subsubsection{Setup}

\paragraph*{Data.} We use the English portion of Granary as our source training dataset $\mathcal{D}_{\text{source}}$, and LibriSpeech, CommonVoice, and TED-LIUM as our target domains $  \mathcal{D}_{\text{target}}$. The test sets of the target datasets are kept untouched; only their validation sets are used for subset selection. We apply Whisper normalizer\footnote{\url{https://github.com/openai/whisper}} to all transcripts and use the same $8$k-token word-piece tokenizer used in~\citep{gu2025omni}. The audio is represented as $80$-dimensional log-mel filterbank features, extracted using a $25$ ms sliding window with a $10$ ms stride.

\paragraph*{Embeddings.} 
We extract speech and text embeddings from pretrained models. Speaker representations are extracted from an MFA-Conformer speaker recognition model~\citep{aldeneh2025speaker}. We use activations from the $3072$-dimensional penultimate layer. These embeddings are subsequently projected to $256$-dimensional space using Gaussian random projection~\citep{bingham2001random}.\footnote{We quantify cosine preservation after random projection by computing the correlation of pairwise cosine similarities over $100$k randomly sampled pairs. We obtain a correlation of $0.96$ and $0.98$ for speaker and WavLM, respectively, indicating that Gaussian random projection preserves relative geometry.} We use a pretrained \texttt{WavLM Base+}\footnote{\url{https://github.com/microsoft/unilm/blob/master/wavlm/README.md}} model with $768$-dimensional embeddings to extract frame-level speech representations that primarily capture phonetic and acoustic information. These embeddings are mean-pooled over time to obtain a single utterance-level representation, which is then projected to $256$ dimensions using Gaussian random projection. Finally, to capture semantic information, we use SBERT with the \texttt{all-MiniLM-L12-v2} model to extract $384$-dimensional text-based sentence embeddings.\footnote{\url{https://sbert.net/docs/sentence_transformer/pretrained_models.html}} These embeddings are used at their original dimensionality and are not projected to a lower dimension, as they are already compact.


\vspace{-0.3cm}
\paragraph*{Model Architecture.} We implement two variants of the Conformer model for our experiments: Conformer-Small and Conformer-Large. Conformer-Small ($9$M parameters) consists of a convolutional encoder with $1$D convolution (kernel size $7$, stride $3$, $288$ filters) followed by $16$ Conformer blocks with an embedding dimension of $144$, $4$ attention heads, $576$ MLP hidden units, and dropout of $0.1$. Conformer-Large ($107$M parameters) follows the same architecture but with larger dimensions: $1536$ convolutional filters, an embedding dimension of $512$, and $2048$ MLP hidden units, while maintaining $16$ Conformer blocks, $4$ attention heads, and a dropout rate of $0.1$.

\vspace{-0.3cm}
\paragraph*{Training Recipe.} Models are trained for $500$k steps on $8\times$A$100$ ($80$GB) GPUs using AdamW optimizer and CTC loss. Each GPU processes batches of up to $500$ seconds of speech. We apply gradient clipping with a maximum norm of $0.5$. 
The input features to the models are $80$-dimensional log-mel filterbanks computed with a $25$ ms window and a $10$ ms stride. 
We use SpecAugment~\citep{park2019specaugment} after $10$k steps with two frequency masks (maximum width $30$) and ten time masks (maximum width $50$, maximum time-masking ratio $0.1$). The learning rate is linearly warmed up to $0.002$ over the first $10$k steps. Beginning at step $200$k, we apply a cosine decay schedule, which is updated every $50$k steps.

\vspace{-0.3cm}
\paragraph*{Data Selection.} 
We set the MMR trade-off parameter to $\lambda=0.7$, and ablate its effect in a later experiment. To scale MMR to large datasets, we implement several modifications to improve its efficiency: a batched selection procedure, candidate pre-filtering, and limiting the number of target embeddings via $k$-means clustering ($k=200$). The detailed selection procedure is summarized in Algorithm~\ref{alg:greedy-mmr}.

\subsubsection{Results}

We present our experimental results in two parts. Tables~\ref{tab:main_results_1} and~\ref{tab:main_results_2} establish the value of embedding-based data selection by comparing models trained on the full Granary corpus, random subsets, and MMR-selected subsets across all three target domains. Table~\ref{tab:main_results_3} then examines multi-dataset selection strategies. We address each of these comparisons in turn through the following questions.

\begin{table*}[t]
\centering
\caption{WER (\%) of Conformer models trained on different source datasets. The models are evaluated on LibriSpeech (clean/other), CommonVoice (CV), and TED-LIUM test sets. Results are reported for both Conformer-Large and Conformer-Small architectures. ``Granary (Random 5\%)'' denotes training on a randomly selected 5\% subset of Granary, while ``Granary (Full)'' uses the complete dataset. Lower WER indicates better performance. Best results are shown in \textbf{bold}, and second-best results are \underline{underlined}.}
\label{tab:main_results_1}
\setlength{\tabcolsep}{6pt}
\begin{tabular}{llclccc}
\toprule
\textbf{Model} & \textbf{Training Data}  &
\textbf{LS-clean} & \textbf{LS-other} & \textbf{CV} & \textbf{TED-LIUM} \\
\midrule

\multirow{6}{*}{Conformer-Small}
 & LS-960                 & \textbf{5.2}  & \textbf{12.2} & 45.4 & 18.9 \\
 & CommonVoice              & 27.8 & 39.4 & \textbf{32.7} & 43.8 \\
 & TED-LIUM                 & 13.5 & 24.5 & 41.3 & \textbf{9.4} \\
  \cline{2-6}

 & Granary (Full)           & \underline{12.5} & \underline{23.2} & \underline{36.6} & 11.0 \\
 & Granary (Random 5\%)    & \underline{12.5} & 23.5 & 37.1 & \underline{10.7} \\
  & \quad \textbf{$\Delta \%$ (Full → Random 5\%) } &
 +0.0\% & +1.3\% & +1.4\% & -2.7\% \\

\midrule\midrule

\multirow{6}{*}{Conformer-Large}
 & LS-960                & \textbf{3.2}  & \textbf{8.0}  & 40.4 & 14.9 \\
 & CommonVoice             & 12.6 & 21.3 & \textbf{19.3} & 21.4 \\
 & TED-LIUM               & 11.8 & 22.5 & 40.2 & 7.8 \\
 \cline{2-6}
 & Granary (Full)           & \underline{6.7}  & \underline{14.3} & \underline{25.4} & \textbf{6.5} \\
 & Granary (Random 5\%)        & 7.1  & 14.9 & 26.2 & \underline{6.9} \\
  & \quad \textbf{$\Delta \%$ (Full → Random 5\%)} &
 +6.0\% & +4.2\% & +3.1\% & +6.2\% \\
\bottomrule
\end{tabular}
\end{table*}


\begin{table*}[t]
\centering
\caption{WER (\%) of Conformer-Small models trained on Granary using different $5\%$ data selection strategies. Results are reported on LibriSpeech (clean/other), CommonVoice (CV), and TED-LIUM test sets. ``Fusion'' indicates the use of a multi-embedding subset selection strategy that combines all three embedding types, as described in Section~\ref{subsec:multi-embedding}. Lower WER indicates better performance. Best results are shown in \textbf{bold}, and second-best results are \underline{underlined}.}
\label{tab:main_results_2}
\setlength{\tabcolsep}{8pt} 
\resizebox{\columnwidth}{!}{
\begin{tabular}{lllc cccc}
\toprule
\textbf{Subset} & \textbf{Model} & \textbf{Approach} & \textbf{Embedding} &
\textbf{LS-clean} & \textbf{LS-other} & \textbf{CV} & \textbf{TED-LIUM} \\
\midrule
Full & Conformer-Small & \multicolumn{1}{c}{--} & -- & 12.5 & 23.2 & 36.6 & 11.0 \\
\midrule
\multirow{8}{*}{5\%}
 &  \multirow{8}{*}{Conformer-Small}  & Random (Baseline) & -- & 12.5 & 23.5 & 37.1 & 10.7 \\
 &  & \quad$+$ Fine-Tune & -- & 14.1 & 25.2 & \underline{33.6} & 17.8 \\
 &  & Duration (Baseline)& -- & 13.2 & 23.8 & 36.5 & 10.8 \\
 \cmidrule(lr){3-8}
 &  & \multirow{4}{*}{MMR} & Speaker & 10.8 & 20.9 & 34.4 & 9.6 \\
 &   & & WavLM & 10.3 & 20.6 & \underline{33.6} & \textbf{9.2} \\
 &   & & SBERT & \underline{8.9} & \underline{18.4} & 35.7 & 9.9 \\
 &   & & Fusion & \textbf{7.9} & \textbf{17.2} & 34.0 & \underline{9.4} \\
 &  & \quad$+$ Fine-Tune & Fusion & 10.9 & 21.0 & \textbf{31.8} & 16.3 \\
\midrule
\multicolumn{4}{l}{$\Delta \%$ (\textbf{Random → Fusion)}}  &   -36.8\% & -26.8\% & -8.4\% & -12.1\% \\
\midrule
\midrule
Full & Conformer-Large & -- & -- & 6.7  & 14.3 & 25.4 & 6.5 \\
5\% & Conformer-Large & MMR & Fusion & 4.4 & 10.7 & 23.6 & 5.9 \\
\bottomrule
\end{tabular}
}
\vspace{0.5cm}
\end{table*}


\begin{table*}[t]
\centering
\caption{WER (\%) comparing single-dataset and multi-dataset subset selection strategies using different embedding types. Single-dataset selection targets each evaluation domain independently, while multi-dataset strategies select a unified 5\% subset that targets all domains simultaneously, using either average (Avg.) or maximum (Max) aggregation of relevance scores (Section~\ref{subsec:multi_data}). Models are Conformer-Small trained on 5\% selected from Granary. Lower WER indicates better performance. Best results are shown in \textbf{bold}, and second-best results are \underline{underlined}.}
\label{tab:main_results_3}
\setlength{\tabcolsep}{8pt} 
\begin{tabular}{llc cccc}
\toprule
\textbf{Subset} & \textbf{Approach} & \textbf{Embedding} &
\textbf{LS-clean} & \textbf{LS-other} & \textbf{CV} & \textbf{TED-LIUM} \\
\midrule
Full & -- & -- & 12.5 & 23.2 & 36.6 & 11.0 \\
\midrule

\multirow{9}{*}{5\%} 
& Single-dataset & \multirow{3}{*}{Speaker} & \textbf{10.8} & \textbf{20.9} & \underline{34.4} & \underline{\textbf{9.6}} \\
 & Multi-dataset (Avg.) &                   & \underline{11.8} & \underline{21.9} & 37.9 & \underline{\textbf{9.6}} \\
 & Multi-dataset (Max) &                    & 12.6 & 22.1 & \textbf{33.8} & 10.3 \\
 \cmidrule(lr){2-7}
 & Single-dataset & \multirow{3}{*}{WavLM} & \textbf{10.3} & \textbf{20.6} & \textbf{33.6} & \textbf{9.2} \\
 & Multi-dataset (Avg.) &                  & 10.8 & \underline{20.8} & 37.9 & \underline{9.8} \\
 & Multi-dataset (Max) &                   & \underline{10.7} & 21.1 & \underline{36.7} & \underline{9.8} \\
 \cmidrule(lr){2-7}
 & Single-dataset & \multirow{3}{*}{SBERT} & \textbf{8.9} & \textbf{18.4} & \underline{35.7} & \textbf{9.9} \\
 & Multi-dataset (Avg.) &                  & \underline{9.6} & \underline{18.9} & 37.0 & \underline{11.1} \\
 & Multi-dataset (Max) &                   & 10.2 & 20.4 & \textbf{35.4} & 11.4 \\
\bottomrule
\end{tabular}
\end{table*}

\begin{table*}[t]
\centering
\caption{WER (\%) of Conformer models trained on different randomly sampled subsets of Granary (1–100\%) evaluated on LibriSpeech (clean/other), CommonVoice, and TED-LIUM test sets. Lower WER indicates better performance.}
\label{tab:subset_fraction}
\setlength{\tabcolsep}{6pt}
\begin{tabular}{llcccc}
\toprule
\textbf{Model} & \textbf{Subset ($\%$)} & \textbf{LS-clean} & \textbf{LS-other} & \textbf{CV} & \textbf{TED-LIUM} \\
\midrule
\multirow{6}{*}{Conformer-Small}
 & 100 & 12.5 & 23.2 & 36.6 & 11.0 \\
 & 50  & 12.1 & 22.7 & 35.7 & 10.7 \\
 & 20  & 12.3 & 22.8 & 36.5 & 11.0 \\
 & 10  & 12.5 & 23.1 & 36.7 & 10.8 \\
 & 5   & 12.5 & 23.1 & 36.7 & 11.2 \\
 & 1   & 12.9 & 24.2 & 38.6 & 11.5 \\
 \midrule
 \multirow{6}{*}{Conformer-Large}
 & 100 & 6.7  & 14.3 & 25.4 & 6.5 \\
 & 50  & 6.8  & 14.3 & 25.6 & 6.4 \\
 & 20  & 6.7  & 14.2 & 25.5 & 6.3 \\
 & 10  & 7.1  & 14.8 & 26.1 & 6.4 \\
 & 5   & 7.0  & 14.9 & 26.0 & 6.5 \\
 & 1   & 8.8  & 17.9 & 30.6 & 7.8 \\
\bottomrule
\end{tabular}
\end{table*}

\vspace{-0.3cm}
\paragraph*{How does the performance of an ASR model trained on the large-scale Granary dataset compare to models trained directly on the training sets of each target domain?} 
Table \ref{tab:main_results_1} shows that training on the large-scale Granary dataset consistently yields stronger cross-domain generalization than training directly on the in-domain data of each target dataset. Models trained exclusively on LibriSpeech or CommonVoice achieve the best performance on their respective domains but exhibit degradation when evaluated on mismatched test sets. Models trained on  TED-LIUM, however, lag behind models trained on the Granary dataset. Training on the full Granary dataset provides a balance between robustness and specialization, even surpassing in-domain training on TED-LIUM.

These results show that large-scale, in-the-wild training data provides a strong foundation for ASR models when robustness across domains is required. However, they also reveal an important challenge: domain mismatch degrades performance. In practice, we often lack sufficient in-domain labeled data to train models, thus we rely on large-scale in-the-wild pseudo-labeled data---which, while abundant, can be poorly matched to target domains. This gap motivates our central question: can strategic data selection bridge the domain mismatch to extract more value from our in-the-wild training data?

\vspace{-0.3cm}
\paragraph*{How does ASR performance degrade when using a smaller model instead of a larger one, and when training on a random $5\%$ subset of Granary instead of the full dataset?} Table \ref{tab:main_results_1} shows that the Conformer-Large model consistently outperforms the Conformer-Small model, both in cross-domain generalization and when trained directly on the in-domain data of each target dataset. The Conformer-Large model exhibits a larger increase in WER when trained on a $5\%$ random subset of Granary compared to the increase exhibited by the Conformer-Small model. This result indicates that the larger model is more sensitive to reductions in the scale of the training data. Similarly, the results suggest that a small model does not have the capacity to model all of the details of a large dataset. Interestingly, down-sampling Granary improves performance on the TED-LIUM dataset for the small model, suggesting that using some samples can mitigate domain mismatch in this target domain.

\vspace{-0.3cm}
\paragraph*{What is the effect of selecting training data based solely on utterance duration rather than embedding-based characteristics?} A baseline we examine the impact of using utterance duration in our selection process to match the distribution of the durations in the target domain. In Table~\ref{tab:main_results_2}, we observe that utilizing utterance duration for down-sampling generally performs slightly worse compared to random down-sampling. However, in the case of CommonVoice, there is a slight improvement over random down-sampling. This result is reasonable, since CommonVoice typically contains shorter utterances, on average, compared to the other datasets.

\begin{wrapfigure}{r}{0.45\textwidth}
  \centering
  \includegraphics[width=0.45\textwidth]{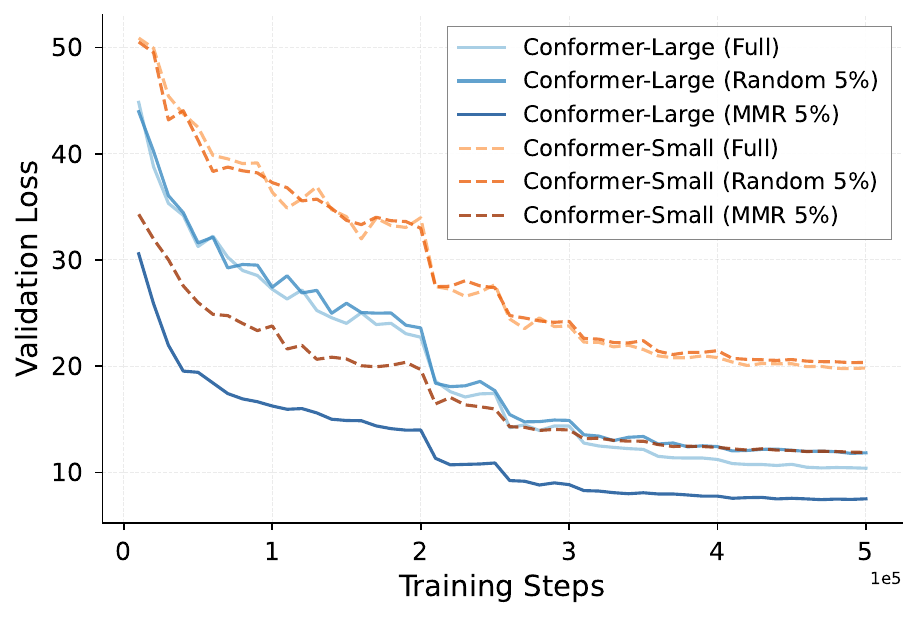}
  \caption{Validation loss curves for Conformer-Large and Conformer-Small models trained on the Granary dataset with the LibriSpeech-clean validation set as a target. Models are compared across three training data conditions: the full dataset, a random $5\%$ subset, and an MMR-selected $5\%$ subset. The MMR selection strategy achieves a lower validation loss than using the full dataset with only $5\%$.}
  \label{fig:learning_curve}
\end{wrapfigure}

\vspace{-0.3cm}
\paragraph*{What is the impact of embedding-based data selection on ASR performance?} 
Table \ref{tab:main_results_2} compares different $5\%$ data selection strategies for training a Conformer-Small model on Granary. Random sampling at $5\%$ yields performance close to using the full dataset, indicating redundancy in Granary and the inability of a small model to leverage the data. In contrast, using an MMR-based selection strategy substantially improves performance over a random down-sampling strategy, with gains that depend on the embedding space used to define relevance and diversity. Speaker and WavLM embeddings lead to consistent reductions in WER across all evaluation sets. SBERT-based selection yields the largest improvements on LibriSpeech but degrades performance on CommonVoice. The proposed multi-embedding fusion achieves the best overall average performance across the four targets, with gains driven primarily by improvements on the LS-clean and LS-other test sets. These results highlight the importance of both embedding choice and cross-domain robustness in data selection. These results also highlight the need for strategic data selection when training on in-the-wild data to maximize performance on a downstream task.
These findings extend to larger models. Table~\ref{tab:main_results_2} shows that Conformer-Large similarly benefits from embedding-based selection. This shows that strategic data selection remains important even as model capacity increases.

\newpage
Figure~\ref{fig:learning_curve} shows the validation loss across three training setups: the full dataset, a randomly selected $5\%$ subset, and an MMR-selected $5\%$ subset. Results are shown for both the large and small model variants. The figure shows that training on the MMR-selected subset consistently yields lower validation loss than training on either a random subset or the full dataset. This trend holds for both model sizes. In addition, the small model trained with MMR achieves validation loss comparable to that of the large model trained on the full dataset or on a random subset.

\vspace{-0.3cm}
\paragraph*{What if we fine-tune our models on the respective validation sets of the test data?} We study whether fine-tuning on the validation set of each target dataset improves performance beyond subset selection alone. Table~\ref{tab:main_results_2} shows the results for two fine-tuning scenarios: fine-tuning a model trained on a randomly sampled $5\%$ subset of Granary, and fine-tuning the MMR Fusion model. In most cases, fine-tuning leads to degraded performance relative to the corresponding non–fine-tuned models, particularly on LibriSpeech and TED-LIUM. This increase in WER can be attributed to overfitting to small validation sets. In contrast, CommonVoice benefits from fine-tuning, with the random $5\%$ model improving from $37.1\%$ to $33.6\%$ WER and the MMR Fusion model improving further to $31.8\%$. These results suggest that, when sufficient validation set is available, further fine-tuning can improve performance beyond data selection alone.

\vspace{-0.3cm}
\paragraph*{What is the effect of using a multi-dataset subset selection strategy on target ASR Performance?} Table \ref{tab:main_results_3} shows that the multi-dataset selection strategies using maximum and mean aggregation methods generally lead to degraded or inconsistent ASR performance across evaluation sets compared to dataset-specific subset selection. For Speaker and WavLM embeddings, both aggregation methods (maximum and mean) increase WER on LibriSpeech (clean and other) relative to their non-merged counterparts, indicating that combining datasets during selection weakens in-domain performance. The impact on CommonVoice and TED-LIUM is mixed: maximum aggregation sometimes improves out-of-domain results (e.g., on CommonVoice), but these gains are not consistent and often come at the cost of LibriSpeech accuracy. For SBERT embeddings, both aggregation strategies underperform targeted selection across all test sets. Overall, these results show that preserving dataset-specific structure during subset selection is more effective than enforcing a unified selection space across multiple datasets.

\subsubsection{Takeaways}
We showed that data selection is an effective strategy to accommodate train/test mismatches when training on large-scale in-the-wild datasets. Using a fixed large corpus, we compare several data selection approaches based on off-the-shelf embeddings, including speaker embeddings, WavLM, and SBERT. Our results show that these methods capture more than simple heuristics such as utterance duration. Moreover, we find that small models are unable to fully exploit the variability in large datasets: Even when selecting only $5\%$ of the data, performance drops only marginally compared to training on the full corpus.

\begin{table}[t]
\centering
\caption{WER (\%) on LibriSpeech (clean/other), CommonVoice (CV), and TED-LIUM test sets for different embeddings and MMR trade-off parameter values ($\lambda$) as described in Equation~\ref{eq2}. Models are Conformer-Small trained on 5\% selected from Granary. Lower WER indicates better performance. Best results are shown in \textbf{bold}, and second-best results are \underline{underlined}.}
\label{tab:mmr_lambda}
\setlength{\tabcolsep}{6pt}
\begin{tabular}{llcccc}
\toprule
\textbf{Embedding} & \textbf{$\lambda$} & \textbf{LS-clean} & \textbf{LS-other} & \textbf{CV} & \textbf{TED-LIUM} \\
\midrule

\multirow{3}{*}{Speaker} 
 & 0.0 & 14.0 & 23.6 & 35.8 & 11.3 \\
 & 0.7 & \underline{10.8} & \underline{20.9} & \underline{34.4} & \underline{9.6} \\
 & 1.0 & \textbf{10.1} & \textbf{20.8} & \textbf{34.3} & \textbf{9.4} \\
\midrule

\multirow{3}{*}{WavLM} 
 & 0.0 & 12.9 & 22.2 & 35.3 & 11.4 \\
 & 0.7 & \underline{10.3} & \textbf{20.6} & \textbf{33.6} & \underline{\textbf{9.2}} \\
 & 1.0 & \textbf{9.6}  & \underline{20.7} & \underline{35.0} & \underline{\textbf{9.2}} \\
\midrule

\multirow{3}{*}{SBERT} 
 & 0.0 & 10.6 & 19.7 & \underline{\textbf{35.7}} & 10.5 \\
 & 0.7 & \textbf{8.9}  & \textbf{18.4} & \underline{\textbf{35.7}} & \textbf{9.9} \\
 & 1.0 & \underline{9.3}  & \underline{18.6} & 36.5 & \underline{10.1} \\
\bottomrule
\end{tabular}
\end{table}

\subsection{Ablations}

The goal of our next set of experiments is to run additional ablations to understand better the sources of the performance gains provided by data selection methods.

\vspace{-0.3cm}
\paragraph*{How does ASR performance vary as we increase the fraction of randomly selected training data from Granary?} 
Table \ref{tab:subset_fraction} examines the effect of training data scale for both Conformer-Large and Conformer-Small models. For the Conformer-Large model, the mean performance across the four datasets remains stable when reducing the training data from $100\%$ down to $5\%$, with a mean degradation of $3\%$. This result suggests that the training dataset contains substantial redundancy for our target datasets, and we do not lose much performance in the target domain by using only $5\%$ of the original training data. In contrast, the degradation is more noticeable when we use $1\%$ of the data. The Conformer-Small model shows lower sensitivity to data reduction than the Conformer-Large model, indicating a lower capacity to compensate for reduced data coverage.

\vspace{-0.3cm}
\paragraph*{How does the relevance–diversity trade-off parameter ($\lambda$) affect ASR performance?} Table~\ref{tab:mmr_lambda} analyzes the effect of the MMR trade-off parameter $\lambda$ from Equation~\ref{eq2} across different embedding types. Setting $\lambda > 0$ generally yields the strongest results; although the optimal setting varies slightly across embeddings and evaluation domains. Speaker and WavLM embeddings benefit most from larger $\lambda$ values. In contrast, SBERT achieves its best performance at $\lambda = 0.7$, with performance degrading at $\lambda = 1.0$, suggesting that semantic embeddings are more sensitive to over-emphasizing relevance at the expense of diversity. Overall, these results highlight the importance of tuning $\lambda$ and demonstrate that the relevance-diversity trade-off depends on the choice of embedding representation.

\vspace{-0.3cm}
\paragraph*{Do the embeddings capture complementary information?}  To quantify the degree of shared information between speaker, WavLM, and SBERT embeddings, we analyze cross-embedding predictability using a logistic regression-based probe. Specifically, we first cluster one embedding space and treat the resulting cluster assignments as pseudo-labels. We then train a logistic regression classifier to predict these pseudo-labels given the other embedding space. If the clusters of one representation can be accurately predicted from the other representations, then this result indicates that the two embeddings encode similar factors of variation. Conversely, if the clusters cannot be accurately predicted, then the target embedding captures information that is not encoded by the source embedding (i.e., complementary representations). 

The results in Table~\ref{tab:cluster_predictability} show the predictability of one embedding space from another using logistic regression on 100-cluster assignments. As expected, each embedding predicts its own clusters most accurately. Cross-embedding prediction accuracies are lower: WavLM modestly predicts speaker clusters (21.4\%) and poorly predicts SBERT clusters (5.9\%). Speaker and SBERT embeddings show similarly low cross-predictability. These results indicate that the three embeddings capture largely complementary information.

\begin{table}[t]
\centering
\caption{Cross-embedding predictability using logistic regression on $100$-cluster assignments. Accuracy is reported in percentage. Accuracy for a random classifier is around 1\%. Best results are \textbf{boldfaced}.}
\label{tab:cluster_predictability}
\begin{tabular}{l l c}
\toprule
\textbf{Source} & \textbf{Target} & \textbf{Accuracy (\%)} \\
\midrule
\multirow{3}{*}{WavLM} & WavLM   & \textbf{76.6} \\
   & Speaker & 21.3 \\
   & SBERT   & 5.9  \\
\midrule
\multirow{3}{*}{Speaker} & WavLM   & 18.8 \\
 & Speaker & \textbf{71.0} \\
 & SBERT   & 3.7  \\
\midrule
\multirow{3}{*}{SBERT}   & WavLM   & 4.6  \\
   & Speaker & 3.9  \\
   & SBERT   & \textbf{87.9} \\
\bottomrule
\end{tabular}
\end{table}
\vspace{-0.1cm}


\section{Discussion \& Conclusion}
We studied ASR training data selection when the target set comes from a different distribution. \textbf{We showed that using an embedding-based MMR approach to strategically select just $\textbf{5\%}$ of the $\textbf{100}$k-hour Granary corpus, we exceed the performance achieved when training on the full dataset with production-size models ($\textbf{10}$-$\textbf{100}$M parameters).} Guiding the selection process with any embedding type (speaker, phonetic, and semantic) provided improvements over training on the full data or using a random selection approach. The use of a multi-embedding approach yielded the best average performance across the evaluated test sets; however, the improvements were mainly driven by gains on LibriSpeech. Our results also showed that the relevance-diversity trade-off depends on the choice of embedding representation. 
Merging target datasets via maximum or mean aggregation consistently underperformed dataset-specific selection, suggesting that different domains impose conflicting selection attributes.
For practitioners deploying resource-constrained specialist models, our results suggest prioritizing data relevance and diversity over sheer scale.

Despite the benefits shown and optimizations explored, one limitation worth noting is that our greedy MMR procedure is computationally expensive. Another limitation is that the reliance on the pseudo-labeled Granary data may introduce label noise. 

\section{Generative AI Use Disclosure}
Generative AI tools were used for language editing, including improving readability, grammar, and transitions between sentences. AI tools were also used to assist with writing some functions for our code. The tools were not used to generate scientific content, experimental results, analysis, or conclusions. All authors take full responsibility for the content of the manuscript.

\section{Acknowledgement}
We thank Andrew Silva, Li-Wei Chen, Amrit Romana, Shiladitya Dutta, Federico Villalpando Paez, Navdeep Jaitly, and Russ Webb for their feedback. We thank Jerremy Holland and Samy Bengio for their support.


\bibliographystyle{apalike}
\bibliography{mybib}
\end{document}